\documentclass[a4paper,12pt]{article}

\usepackage[utf8]{inputenc}
\usepackage[T1]{fontenc}
\usepackage{amsmath, amssymb, amsxtra}
\usepackage{nccmath, icomma, xspace, bbold}
\usepackage[mathic]{mathtools}
\usepackage[version=3]{mhchem}
\usepackage{siunitx}
\DeclareSIUnit\krec{\ensuremath{\textit{k}_\text{rec}}}
\DeclareSIUnit\Erec{\ensuremath{\textit{E}_\text{rec}}}
\usepackage[short]{datetime}
\usepackage[normalem]{ulem}
\usepackage{color}

\usepackage[%
	left = 2cm,%
	right = 2cm,%
	top = 2cm,%
	bottom = 2.5cm,%
	footskip = 1.5cm,%
	nohead]
	{geometry}
\usepackage{nccparskip}
\SetParskip{4pt}
\usepackage{setspace}
\usepackage{enumitem}

\usepackage[british]{babel}
\usepackage{csquotes}

\usepackage{ctable}
\usepackage{caption}%
\DeclareCaptionLabelSeparator{vertline}{~|~}%
\usepackage[%
	backend=bibtex,%
	style=nature,%
	url=false,%
	doi=false]
	{biblatex}
\bibliography{HOM.bib}
\AtEveryBibitem{%
	\clearfield{day}%
	\clearfield{endday}%
	\clearfield{month}%
	\clearfield{endmonth}%
	\clearfield{issn}%
	\clearfield{isbn}%
	\clearfield{issue}%
	}

\usepackage{authblk}

\usepackage{xifthen}
\newcommand{\ket}[2][]{ %
	\ifthenelse{\isempty{#1}}%
	{\ensuremath{\xspace\left\vert #2 \right\rangle\xspace}}%
	{\ensuremath{\xspace\left\vert #2 \right\rangle_{\! #1}\xspace}}
	}
\newcommand{\bra}[2][]{ %
	\ifthenelse{\isempty{#1}}%
	{\ensuremath{\xspace\left\langle #2 \right\vert\xspace}}
	{\ensuremath{\xspace\prescript{}{#1}{\!\left\langle #2 \right\vert\xspace}}}
	}
\newcommand{\braket}[3][]{ %
	\ifthenelse{\isempty{#1}}%
	{\ensuremath{\xspace\left\langle #2 \left\vert\right. #3 \right\rangle\xspace}}
	{\ensuremath{\xspace\left\langle #2 \left\vert\right. #3 \right\rangle_{\! #1}\xspace}}
	}
\newcommand{\avg}[2][]{ %
	\ifthenelse{\isempty{#1}}%
	{\ensuremath{\xspace\langle #2 \rangle\xspace}}%
	{\ensuremath{\xspace\langle #2 \rangle_{\! #1}\xspace}}
	}
\newcommand{\Gtwo}{\ensuremath{G^{(2)}}\xspace}
\newcommand{\krec}{\ensuremath{k_\text{rec}}\xspace}
\newcommand{\Erec}{\ensuremath{E_\text{rec}}\xspace}

\DeclareMathOperator{\Real}{Re}

\title{\bfseries An atomic Hong–Ou–Mandel experiment}
\author{R. Lopes, A. Imanaliev, A. Aspect, M. Cheneau, D. Boiron, C. I. Westbrook}
\affil{Laboratoire Charles Fabry\\Institut d'Optique Graduate School\,–\,CNRS\,–\,Université Paris Sud,\\2 avenue Augustin Fresnel, 91127 Palaiseau, France}

\date{\today}
\begin{document}
\maketitle
\begin{doublespace}

% -------------------------
% Summary abstract
% -------------------------
{\bfseries
Quantum mechanics is a very successful and still intriguing theory, introducing two major counter-intuitive concepts.
\emph{Wave-particle duality} means that objects normally described as particles, such as electrons, can also behave as waves, while entities primarily described as waves, such as light, can also behave as particles.
This revolutionary idea nevertheless relies on notions borrowed from classical physics, either waves or particles evolving in our ordinary space-time.
By contrast, \emph{entanglement} leads to interferences between the amplitudes of multi-particle states, which happen in an abstract mathematical space and have no classical counterpart.
This fundamental feature has been strikingly demonstrated by the violation of Bell's inequalities \supercite{Bell1964, Aspect1999, Giustina2013, Christensen2013}.
There is, however, a conceptually simpler situation in which the interference between two-particle amplitudes entails a behaviour impossible to describe by any classical model.
It was realised in the Hong, Ou and Mandel (HOM) experiment \supercite{Hong1987}, in which two photons arriving simultaneously in the input channels of a beam-splitter always emerge together in one of the output channels.
In this letter, we report on the realisation, with atoms, of a HOM experiment closely following the original protocol.
This opens the prospect of testing Bell's inequalities involving mechanical observables of massive particles, such as momentum, using methods inspired by quantum optics \supercite{Rarity1990, Lewis-Swan2014b}, with an eye on theories of the quantum-to-classical transition \supercite{Penrose1998, Zurek2003, Schlosshauer2005, Leggett2007}.
Our work also demonstrates a new way to produce and benchmark twin-atom pairs \supercite{Bucker2011, Kaufman2014} that may be of interest for quantum information processing \supercite{Nielsen2000} and quantum simulation \supercite{Kitagawa2011}.
}

\begin{figure}[!h]
	\centering
	\includegraphics[width=0.7\columnwidth]{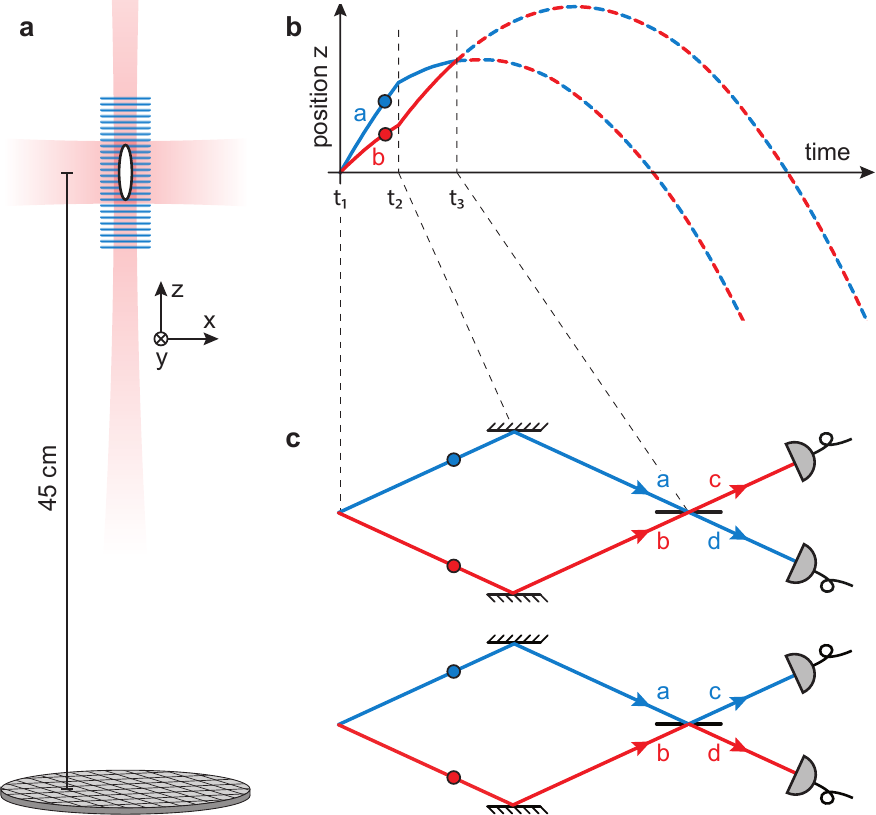}
	\caption{{\bf Schematic of the experiment.} 
	\textbf{a,}
	A Bose–Einstein condensate (BEC, white oval) of metastable Helium 4 atoms is trapped in an elongated optical trap (red shaded area).
	A moving optical lattice, here depicted in blue, is superimposed on the BEC and triggers the scattering of atom pairs along the $z$-axis.
	When this lattice and the trap are switched off, the atoms fall towards a micro-channel plate detector located \SI{45}{cm} below the initial position of the BEC (drawing not to scale).
	\textbf{b,}
	The time diagram shows the evolution of the twin atom vertical coordinates (blue and red lines).
	Between  $t_1$ and $t_2$,  $t_2$ and $t_3$, and after $t_3$, the atoms move under the sole influence of gravity (drawing not to scale).
	At $t_2$, the twin atom velocities are swapped using Bragg diffraction on an optical lattice.
	At time $t_3$, when the atomic trajectories cross again, the same lattice is applied for half the amount of time in order to realise a beam-splitter.
	\textbf{c,}
	In the centre-of-mass frame of reference, the trajectories of the atoms resemble those of the photons in the Hong–Ou–Mandel experiment.
	A joint detection arises either when both atoms are transmitted through the beam-slitter (upper panel) or when both are reflected (lower panel).
	If the two particles are indistinguishable, these processes end in the same final quantum state and the probability of joint detection results from the addition of their amplitudes.
	For bosons these amplitudes have same modulus but opposite signs, thus their sum vanishes and so also the probability of joint detection.
	}
	\label{fig:1}
\end{figure}

A pair of entangled particles is described by a state vector that cannot be factored as a product of two state vectors associated with each particle.
Although entanglement does not require that the two particles be identical \supercite{Aspect1999}, it arises naturally in systems of indistinguishable particles due to the symmetrisation of the state.
A remarkable illustration is the HOM experiment, in which two photons enter in the two input channels of a beam-splitter and one measures the correlation between the signals produced by photon counters placed at the two output channels. 
A joint detection at these detectors arises from two possible processes: either both photons are transmitted by the beam-splitter or both are reflected (Fig.~\ref{fig:1}{c}).
If the two photons are indistinguishable, both processes lead to the same final quantum state and the probability of joint detection results from the addition of their amplitudes.
Because of elementary properties of the beam-splitter, these amplitudes have same modulus but opposite signs, thus their sum vanishes and so also the probability of joint detection (Refs.~\cite{Ou2007, Grynberg2010} and Methods).
In fact, to be fully indistinguishable, not only must the photons have the same energy and polarisation, but their final spatio-temporal modes must be identical.
In the HOM experiment, it means that the two photons enter the beam-splitter in modes that are the exact images of each other.
As a result, when measured as a function of the delay between the arrival times of the photons on the beam-splitter, the correlation exhibits the celebrated \enquote{HOM dip}, ideally going to zero at null delay.

In this letter, we describe an experiment equivalent in all important respects to the HOM experiment, but performed with bosonic atoms instead of photons.
We produce freely propagating twin beams of metastable Helium 4 atoms \supercite{Bonneau2013}, which we then reflect and overlap on a beam-splitter using Bragg scattering on an optical lattice (Ref.~\cite{Cronin2009} and Fig.~\ref{fig:1}).
The photon counters after the beam-splitter are replaced by a time-resolved, multi-pixel atom-counting detector \supercite{Schellekens2005}, which enables the measurement of intensity correlations between the atom beams in well defined spatial and spectral regions.
The temporal overlap between the atoms can be continuously tuned by changing the moment when the atomic beam-splitter is applied.
We observe the HOM dip when the atoms simultaneously pass through the beam-splitter.
Such a correlation has no explanation in terms of classical particles.
In addition, a quantitative analysis of the visibility of the dip also rules out any interpretation in terms of single-particle matter waves. Our observation must instead result from a quantum interference between multi-particle amplitudes.

%-------------------------------
% Description of the experiment
%-------------------------------

\begin{figure}[!h]
	\centering
	\includegraphics[width=0.7\columnwidth]{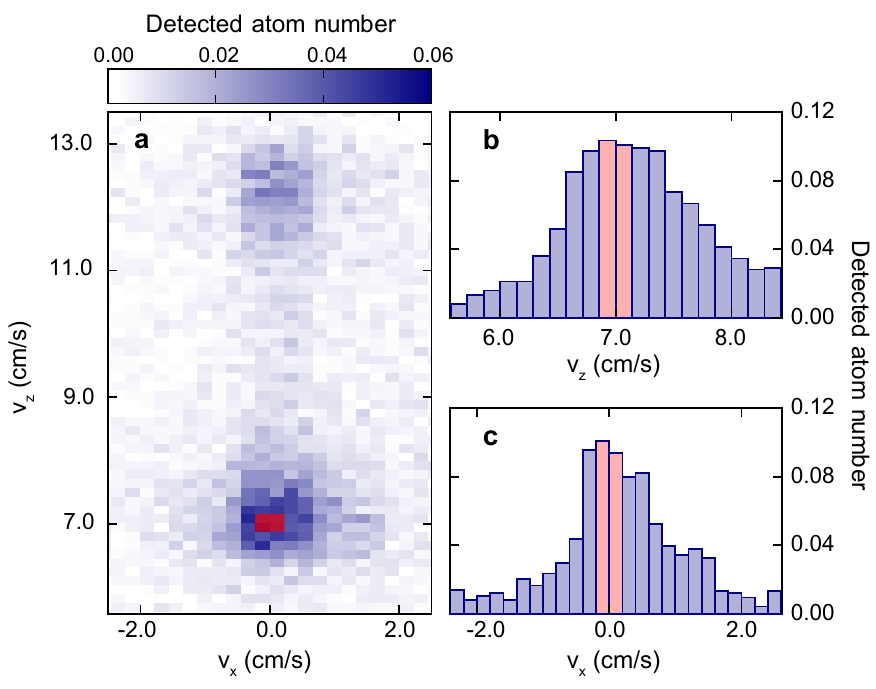}
	\caption{{\bf Velocity distribution of the twin atoms.}
	\textbf{a,}
	Two-dimensional velocity distribution of the twin atom beams emitted by the source.
	The red shaded area, drawn here only for the lower beam, labelled $b$ in Fig.~\ref{fig:1}{b and c}, shows the integration volume used for computing the correlation function displayed in Fig.~\ref{fig:3}.
	The distribution corresponds to an average over about \num{1100} measurements and is not corrected for the limited detection efficiency.
	The velocities are given relative to the laboratory frame of reference.
	The size of each pixel is \SI{0.25}{\cm \per \s} in the transverse directions ($x$ and $y$) and \SI{0.15}{\cm \per \s} in the longitudinal ($z$) direction and an integration over 2 pixels is performed along the $y$ direction.  
	\textbf{b,c,}
	Cuts of the two-dimensional velocity distribution through the centre of the lower beam along the longitudinal (\textbf{b}) and transverse (\textbf{c}) directions.
	The data points result from the average over 2 pixels along the direction perpendicular to the cut.
	The full width at half-maximum of the distribution, obtained from a Gaussian fit, is about $\SI{1.4}{\cm \per \s}$ along both the longitudinal and transverse directions. The red shaded area again shows the integration volume.
	}
	\label{fig:2}
\end{figure}

Our experiment starts by producing a Bose–Einstein condensate (BEC) of metastable Helium 4 atoms in the $2\prescript{3}{}{\text S}_1, m=1$ internal state.
The BEC contains \SIrange{5}{6e4}{atoms} and is confined in an elliptical optical trap with its long axis along the vertical ($z$) direction (Fig.~\ref{fig:1}{a}).
The atomic cloud has radii of 58 and \SI{5}{\um} along the longitudinal and transverse ($\perp$) directions, respectively.
A moving optical lattice, superimposed on the BEC for \SI{300}{\us}, induces the scattering of atom pairs (hereafter referred to as twin atoms) in the longitudinal direction through a process analogous to spontaneous four wave mixing (Refs.~\cite{Hilligsoe2005, Campbell2006, Bonneau2013} and Methods).
One beam, labelled $a$, has a free-space velocity $v_{z} \simeq \SI{12.1}{\cm \per \s}$ in the laboratory frame of reference and the other beam, labelled $b$, has a velocity $v_{z} \simeq \SI{7.0}{\cm \per \s}$ (Fig.~\ref{fig:1}{b,c}).
The twin atom beams clearly appear in the velocity distribution of the atoms, which is displayed in Fig.~\ref{fig:2}.
The visible difference in population between the beams probably results from secondary scattering processes in the optical lattice, leading to the decay over time of the quasi-momentum states \supercite{Bonneau2013}.
After the optical lattice has been switched off (time $t_1$), the twin atoms propagate in the optical trap for \SI{200}{\us}.
At this moment, the trap itself is switched off and the atoms are transferred to the magnetically insensitive $m=0$ internal state by a two-photon Raman transition (Methods).

From here on, the atoms evolve under the influence of gravity and continue to move apart (Fig.~\ref{fig:1}{b}).
At time $t_2 = t_1 + \SI{500}{\us}$, we deflect the beams using Bragg diffraction on a second optical lattice, so as to make them converge.
In the centre-of-mass frame of reference, this deflection reduces to a simple specular reflection (Fig.~\ref{fig:1}{c} and Methods).
At time $t_3 \simeq 2t_2-t_1$, we apply the same diffraction lattice for half the amount of time in order to realise a beam-splitting operation on the crossing atom beams.
Changing the time $t_3$ allows us to tune the degree of temporal overlap between the twin atoms.
Fig.~\ref{fig:1}{c} shows the atomic trajectories in the centre-of-mass frame of reference and reveals the close analogy with a photonic HOM experiment.
The atoms end their fall on a micro-channel plate detector located \SI{45}{cm} below the position of the initial BEC and we record the time and transverse position of each atomic impact with a detection efficiency $\eta \sim \SI{25}{\percent}$ (Methods).
The time of flight to the detector is approximately \SI{300}{\ms}, long enough that the recorded signal yields the three components of the atomic velocity.
By collecting data from several hundred repetitions of the experiment under the same conditions, we are able to reconstruct all desired atom number correlations within variable integration volumes of extent $\Delta v_{z} \times \Delta v_{\perp}^2$.
These volumes play a similar role to that of the spatial and spectral filters in the HOM experiment and can be adjusted to erase the information that could allow tracing back the origin of a detected particle to one of the input channels.

%---------
% HOM dip
%---------

The HOM dip should appear in the cross-correlation between the detection signals in the output channels of the beam-splitter (Ref.~\cite{Ou2007} and Methods):
\begin{equation}
	\Gtwo_{cd} = \Big( \frac{\eta}{\Delta v_{z} \Delta v_{\perp}^2} \Big)^2 \iint_{\mathcal V_c \times \mathcal V_d} \avg{ \hat a^\dagger_{\mathbf v_c} \hat a^\dagger_{\mathbf v_d}  \hat a_{\mathbf v_d} \hat a_{\mathbf v_c} }  \; \text{d}^{(3)}\mathbf v_c \, \text{d}^{(3)}\mathbf v_d \; .
	\label{eq:cross-correlation}
\end{equation}
Here, $\hat a_{\mathbf v}$ and $\hat a^\dagger_{\mathbf v}$ denote the annihilation and creation operators of an atom with three-dimensional velocity $\mathbf v$, respectively, $\avg{\cdot}$ stands for the quantum and statistical average and $\mathcal V_{c,d}$ designate the integration volumes centred on the output atom beams $c$ and $d$ (Fig.~\ref{fig:1}{c}).
We have measured this correlation as a function of the duration of propagation $\tau = t_3 - t_2$ between the mirror and the beam-splitter (Fig.~\ref{fig:3}) and for various integration volumes (see supplementary material).
We observe a marked reduction of the correlation when $\tau$ is equal to the duration of propagation from the source to the mirror ($t_3 - t_2 \simeq t_2 - t_1$) and for small enough integration volumes, corresponding to a full overlap of the atomic wave-packets on the beam-splitter.
Fitting the data with an empirical Gaussian profile yields a visibility:
\begin{equation}
	V = \frac{\max_\tau \Gtwo_{cd}(\tau) - \min_\tau \Gtwo_{cd}(\tau)}{\max_\tau \Gtwo_{cd}(\tau)} = \num{0.65 \pm 0.07} \; ,
\end{equation}
where the number in parenthesis stands for the \SI{68}{\percent} confidence interval.
As we shrink the integration volumes, we observe that the dip visibility first increases and then reaches a saturation value, as is expected when the integration volumes become smaller than the elementary atomic modes.
The data displayed in Fig.~\ref{fig:3} were obtained for $\Delta v_z = \SI{0.3}{\cm \per \s}$ and $\Delta v_\perp = \SI{0.5}{\cm \per\s}$, which maximises the reduction of the correlation while preserving a statistically significant number of detection events (see supplementary material).

The dip in the cross-correlation function cannot be explained in terms of classical particles, for which we would have no correlation at all between the detections in the output channels.
When the atoms are viewed as waves however, demonstrating the quantum origin of the dip  necessitates a deeper analysis.
The reason is that two waves can interfere at a beam-splitter and give rise to an intensity imbalance between the output ports.
If, in addition, the coherence time of the waves is finite, the cross-correlation can display a dip similar to the one observed in our experiment.
But once averaged over the phase difference between the  beams, the visibility is bounded from above and cannot exceed \num{0.5} (Refs.~\cite{Ou1988b, Lewis-Swan2014} and Methods).
In our experiment, this phase difference is randomised by the shot-to-shot fluctuations of the relative phase between the laser beams used for Bragg diffraction (Methods).
Since our measured visibility exceeds the  limit for waves by two standard deviations, we can safely rule out any interpretation of our observation in terms of interference between two \enquote{classical} matter waves or, in other words, between two ordinary wave functions describing each of the two particles separately.

%----------------------------
% Analysis of the visibility
%----------------------------

Two contributions may be responsible for the non-zero value of the correlation function at the centre of the dip: the detected particles may not be fully indistinguishable and the number of particles contained in the integration volume may exceed unity for each beam (Refs.~\cite{Ou1988b, Ou1999} and Methods).
The effect of the atom number distribution can be quantified by measuring the intensity correlations of the twin atom beams upstream of the beam-splitter (Fig.~\ref{fig:1}{c}), which bound the visibility of the dip through the relation:
\begin{equation}
	V_\text{max} = 1 - \frac{\Gtwo_{aa} + \Gtwo_{bb}}{\Gtwo_{aa} + \Gtwo_{bb} + 2 \Gtwo_{ab}} \; ,
	\label{eq:visibility_bound}
\end{equation}
where $\Gtwo_{aa}$, $\Gtwo_{bb}$ and $\Gtwo_{ab}$ are defined according to Eq.~\ref{eq:cross-correlation} (Ref.~\cite{Lewis-Swan2014} and Methods).
Here, one immediately sees that the finite probability of having more than one atom in the input channels will lead to finite values of the auto-correlations $\Gtwo_{aa}$, $\Gtwo_{bb}$ and therefore to a reduced visibility.
We have performed the measurement of these correlations following the same experimental procedure as before, except that we did not apply the mirror and beam-splitter.
We find non-zero values $\Gtwo_{aa} = \num{0.016 \pm 0.005}$, $\Gtwo_{bb} = \num{0.047 \pm 0.009}$, and $\Gtwo_{ab} = 0.048(7)$, yielding $V_\text{max} = \num{0.60 \pm 0.10}$, where the uncertainty is the standard deviation of the statistical ensemble.
Because of the good agreement with the measured value, we conclude that the atom number distribution in the input channels entirely accounts for the visibility of the HOM dip.
For the present experiment we estimate the average number of incident atoms to be \num{0.5 \pm 0.1} in $\mathcal V_a$ and \num{0.8 \pm 0.2} in $\mathcal V_b$, corresponding to a ratio of the probability for having two atoms to that for having one atom of \num{0.25 \pm 0.05} and \num{0.40 \pm 0.10}, respectively (Methods).
Achieving much smaller values is possible, for instance by reducing the pair production rate, but at the cost of much lower counting statistics.

Although multi-particle interferences can be observed with particles emitted or prepared independently \supercite{Beugnon2006, Bocquillon2013,Lang2013,Dubois2013,Kaufman2014}, twin particle sources are at the heart of many protocols for quantum information processing \supercite{Nielsen2000} and quantum simulation \supercite{Kitagawa2011}.
The good visibility of the HOM dip in our experiment demonstrates that our twin atom source produces beams which have highly correlated populations and are well mode matched.
This is an important achievement in itself, which may have the same impact for quantum atom optics as the development of twin photon sources using non-linear crystals had for quantum optics (see for instance Ref.~\cite{Ghosh:1987zl}).

% --------
% Figures
% --------

\clearpage

\newpage

\newpage
\begin{figure}[!h]
	\centering
	\includegraphics[width=0.6\columnwidth]{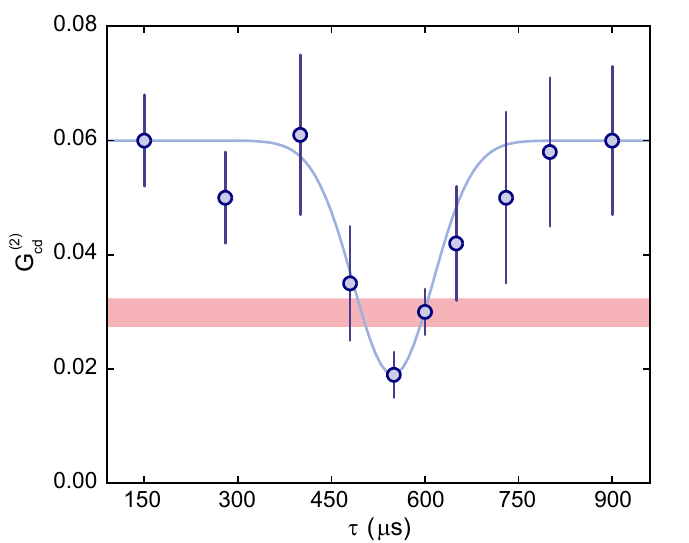}
	\caption{{\bf HOM dip in the cross-correlation function.} 
	 The correlation $\Gtwo_{cd}$ between the output ports of the beam-splitter, defined in Eq.~\ref{eq:cross-correlation}, was measured as a function of the duration of propagation $\tau=t_3- t_2$ between the mirror and the beam splitter.
	 The HOM dip is directly visible as a marked reduction of the correlation when $\tau$ equals the duration of propagation $t_2- t_1 \simeq \SI{500}{\us}$ between the source and the mirror, corresponding to symmetric paths between the source and the beam-splitter, i.e. when one cannot distinguish between the two diagrams of Fig.~\ref{fig:1}{c}.
	 A Gaussian fit (blue line) precisely locates the dip at $\tau = \SI{550 \pm 50}{\us}$, with a full-width at half-maximum of $\SI{150 \pm 40}{\us}$, where the uncertainty corresponds to the \SI{68}{\percent} confidence interval.
	 The measured visibility is $V = \num{0.65 \pm 0.07}$.
	 It is two standard deviations beyond the classical-to-quantum threshold represented by the red shaded area, which takes into account the experimental uncertainty over the background correlation value.
	 Each data point was obtained from an average over about \num{500} to \num{1400} repetitions of the experiment.
	 Error bars denote the standard deviation of the statistical ensemble.
	 The mean detected atom number was constant over the range of values of $\tau$ displayed here (see supplementary material).
	 }
	\label{fig:3}
\end{figure}

\newpage

% ---------------
% End notes
% ---------------

\bigskip
\begin{flushleft}

\textbf{Acknowledgements} We thank Josselin Ruaudel and Marie Bonneau for their contribution to the early steps of the experiment. We also thank Karen Kheruntsyan, Jan Chwedeńczuk and Piotr Deuar for discussions. We acknowledge funding by IFRAF, Triangle de la Physique, Labex PALM, ANR (PROQUP), FCT (scholarship SFRH/BD/74352/2010 to R.L.) and EU (ERC Grant 267775\,–\,QUANTATOP and Marie Curie CIG 618760\,–\,CORENT).

%    \textbf{Author Contributions} All authors contributed extensively to this work.

\textbf{Author Information} Correspondence and requests for materials should be addressed to R.L. (raphael.lopes@institutoptique.fr) or M.C. (marc.cheneau@institutoptique.fr).
\end{flushleft}

% --------
% Methods
% --------

\clearpage  
\newpage
\section*{Methods}

\subsubsection*{Twin atom source}

The twin atom beams result from a scattering process between pairs of atoms from the BEC occurring when the gas is placed in a moving one-dimensional optical lattice.
The experimental set-up has been described  in Ref.~\cite{Bonneau2013}.
The lattice is formed by two laser beams derived from the same source emitting at the wavelength $\lambda = \SI{1064}{\nm}$.
In contrast to our previous work, the axis of the optical lattice was now precisely aligned with the long axis of the optical trap confining the atoms.
The laser beams intersect with an angle of $\theta = \SI{166}{\degree}$, their frequency difference is set to \SI{100.5}{\kHz} and the lattice depth to \SI{0.4}{\Erec}. This constrains the longitudinal wave-vector of the twin atoms to the values $k_{z,a} = \SI{0.75}{\krec}$ and $k_{z,b} = \SI{1.30}{\krec}$ in order to fulfil the conservation of quasi-momentum and energy in the frame co-propagating with the lattice.
Here, $\krec = 2\pi \sin(\theta/2)/\lambda$ is the recoil wave-vector along the longitudinal axis gained upon absorption of a photon from a lattice laser and $\Erec = \hbar^2 \krec^2/2m$ is the associated kinetic energy, with $\hbar$ the reduced Planck constant and $m = \SI{6.64e-27}{\kg}$ the mass of an Helium 4 atom.
The observed velocities of the twin atom beams coincide with the expected values above, using the relation $v = \hbar k/m$.
The optical lattice is turned on and off adiabatically so as to avoid diffraction of the atoms during this phase of the experiment.

\subsubsection*{Transfer to the magnetically insensitive internal state}

Transfer to the $m=0$ state after the optical trap has been switched off is made necessary by the presence of stray magnetic fields in the vacuum chamber that otherwise would lead to a severe deformation of the atomic distribution during the long free fall.
The transfer is achieved by introducing a two-photon coupling between the $m=1$ state, in which the atoms are initially, and the $m=0$ state using two laser beams derived from a single source emitting at $\SI{1083}{\nm}$ and detuned by \SI{600}{\MHz} from the $2\prescript{3}{}{\text S}_1 $ to $2\prescript{3}{}{\text P}_0$ transition.
The frequency difference of the laser beams is chirped across the two-photon resonance so as to realise an adiabatic fast passage transition (the frequency change is \SI{300}{\kHz} in \SI{300}{\us}).
We have measured the fraction of transferred atoms to be \SI{94}{\percent}. The remaining \SI{6}{\percent} stay in the $m=1$ state and are pushed away from the integration volumes by the stray magnetic field gradients.

\subsubsection*{Atomic mirror and beam-splitter}

The mirror and beam-splitter are both implemented using Bragg scattering on a second optical lattice.
This effect can be seen as a momentum exchange between the atoms and the laser beams forming the lattice, a photon being coherently absorbed from one beam and emitted into the other.
In our experiment, the laser beams forming the lattice have a waist of \SI{1}{\mm} and are detuned by \SI{600}{\MHz} from the $2\prescript{3}{}{\text S}_1$ to $2\prescript{3}{}{\text P}_0$ transition (they are derived from the same source as the beams used for the Raman transfer).
In order to fulfil the Bragg resonance condition for the atom beams, the laser beams are made to intersect at an angle of $\SI{32}{\degree}$ and the frequency of one of the beams is shifted by \SI{53.4}{\kHz}.
In addition to this fixed frequency difference, a frequency chirp is performed to compensate for the acceleration of the atoms during their free fall.
The interaction time between the atoms and the optical lattice was \SI{100}{\us} for the mirror operation ($\pi$-pulse) and \SI{50}{\us} for the beam-splitter operation ($\pi/2$-pulse).
The resonance condition for the momentum state transfer is satisfied by all atoms in the twin beams but only pairs of states with a well defined momentum difference are coupled with each other.
We measured the reflectivity of the mirror and the transmittance of the beam-splitter to be \num{0.95 \pm 0.02} and \num{0.49 \pm 0.02}, respectively. Spontaneous scattering of photons by the atoms was negligible.

\subsubsection*{Detection efficiency}

Our experiment relies on the ability to detect the atoms individually. The detection efficiency is an essential parameter for achieving good signal to noise ratios, although it does not directly influence the visibility of the HOM dip.
Our most recent estimate of the detection efficiency relies on the measurement of the variance of the atom number difference between the twin beams.
For this we use the same procedure as described in Ref.~\cite{Bonneau2013}, but with an integration volume that includes the entire velocity distribution of each beam.
We find a normalised variance of \num{0.75 \pm 0.05}, well below the Poissonian floor.
Since for perfectly correlated twin beams the measured variance would be $1-\eta$, we attribute the lower limit of \SI{25 \pm 5}{\percent} to our detection efficiency.
This value for $\eta$ is a factor of about 2 larger than the lower bound quoted in Ref.~\cite{Jaskula2010}.
The difference is due to the change of method employed for transferring the atoms from the $m=1$ to the $m=0$ state after the optical trap has been switched off.
We previously used  a radio-frequency transfer with roughly \SI{50}{\percent} efficiency whereas the current optical Raman transfer has close to \SI{100}{\percent} efficiency.

\subsubsection*{Distribution of the number of incident atoms}

We have estimated the average number of incident atoms in each input channel of the beams-splitter, $n_a$ and $n_b$, by analysing the distribution of detected atoms in the integration volumes $\mathcal V_a$ and $\mathcal V_b$.
We fitted these distributions by assuming an empirical Poissonian law for the distribution of incident atoms and taking into account the independently calibrated detection efficiency.
The values of $n_a$ and $n_b$ given in the main text are the mean values of the Poissonian distributions that best fit the data.
The probabilities for having one or two atoms in each of the input channels of the beam splitter was obtained from the same analysis.
The uncertainty on these numbers mostly stems from the uncertainty on the detection efficiency.

\subsubsection*{The HOM effect}

The HOM effect appears in the correlator $\avg{ \hat a^\dagger_{\mathbf v_c} \hat a^\dagger_{\mathbf v_d} \hat a_{\mathbf v_d} \hat a_{\mathbf v_c} }$ of Eq.~\ref{eq:cross-correlation}.
The simplest way to calculate such a correlator is to transform the operators and the state vector back in the input space before the beam-splitter and to use the Heisenberg picture.
The transformation matrix between the operators $\hat a_{\mathbf v_c}(t_3)$, $\hat a_{\mathbf v_d}(t_3)$ and $\hat a_{\mathbf v_a}(t_3)$, $\hat a_{\mathbf v_b}(t_3)$ can be worked out from first principles.
For the Bragg beam-splitter, and using a Rabi two-state formalism, we find:
\begin{equation}
	\begin{dcases}
		\hat a_{\mathbf v_c} = \frac{1}{\sqrt 2} \left( i\,e^{i\phi}\,\hat a_{\mathbf v_a} + \hat a_{\mathbf v_b} \right) \; , \\
		\hat a_{\mathbf v_d} = \frac{1}{\sqrt 2} \left( \hat a_{\mathbf v_a} + i\,e^{-i\phi}\,\hat a_{\mathbf v_b} \right) \, ,
	\end{dcases}
\end{equation}
where $\phi$ is the relative phase between the laser beams forming the optical lattice.
In the ideal case of an input state with exactly one atom in each channel, $\ket{1_{\mathbf v_a}, 1_{\mathbf v_b}}$, we therefore obtain:
\begin{align}
	\big\lVert \, \hat a_{\mathbf v_d} \hat a_{\mathbf v_c} \ket{1_{\mathbf v_a}, 1_{\mathbf v_b}} \big\rVert^2
	& = \frac{1}{4} \, \big\lVert \left( i\,e^{i\phi}\,\hat a_{\mathbf v_a}^2 + i\,e^{-i\phi}\,\hat a_{\mathbf v_b}^2 + \hat a_{\mathbf v_a} \hat a_{\mathbf v_b}  + i^2 \, \hat a_{\mathbf v_b} \hat a_{\mathbf v_a} \right) \ket{1_{\mathbf v_a}, 1_{\mathbf v_b}} \big\rVert^2 \\
	& = \frac{1}{4} \, \big\lVert \, 0 + \left( 1+i^2 \right) \ket{0_{\mathbf v_a}, 0_{\mathbf v_b}} \big\rVert^2 \\
	& = 0 \; .
\end{align}
meaning that the probability of joint detection is strictly zero.
The detailed calculation above makes clear that the perfect destructive interference between the two-particle state amplitudes associated with the two diagrams of Fig.~\ref{fig:1}{c} is at the heart of the HOM effect.
By contrast, input states containing more than one atom per channel are transformed into a sum of orthogonal states and the interference can only be partial.
Taking $\ket{2_{\mathbf v_a}, 2_{\mathbf v_b}}$, for instance, yields:
\begin{align}
	\big\lVert \, \hat a_{\mathbf v_d} \hat a_{\mathbf v_c} \ket{2_{\mathbf v_a}, 2_{\mathbf v_b}} \big\rVert^2
	& = \frac{1}{4} \, \big\lVert \left( i\,e^{i\phi}\,\hat a_{\mathbf v_a}^2 + i\,e^{-i\phi}\,\hat a_{\mathbf v_b}^2 + \hat a_{\mathbf v_a} \hat a_{\mathbf v_b}  + i^2 \, \hat a_{\mathbf v_b} \hat a_{\mathbf v_a} \right) \ket{2_{\mathbf v_a}, 2_{\mathbf v_b}} \big\rVert^2 \\
	& = \frac{1}{2} \, \big\lVert \, i\,e^{i\phi} \, \ket{0_{\mathbf v_a}, 2_{\mathbf v_b}} + i\,e^{-i\phi} \, \ket{2_{\mathbf v_a}, 0_{\mathbf v_b}} + \sqrt{2} \left( 1 + i^2 \right) \ket{1_{\mathbf v_a}, 1_{\mathbf v_b}} \big\rVert^2 \\
	& = \frac{1}{2} \, \big\lVert e^{i\phi} \, \ket{0_{\mathbf v_a}, 2_{\mathbf v_b}} + e^{-i\phi} \, \ket{2_{\mathbf v_a}, 0_{\mathbf v_b}} \big\rVert^2 \\
	& = 1 \; .
\end{align}

Finally, we note that losses in one of the incident beams, for instance beam $a$, can be modelled by a fictitious beam-splitter with a transmission coefficient $T$.
In the above calculation, these losses would therefore only manifest by an additional factor $\sqrt T$ in front of every operator $\hat a_{\mathbf v_a}$, leaving unaffected the destructive interference that gives rise to the HOM effect.

\subsubsection*{Visibility of the HOM dip}

A slightly less general form of Eq.~\ref{eq:visibility_bound} has been derived in Ref.~\cite{Lewis-Swan2014} assuming a two-mode squeezed state as an input state.
The same calculation can be performed for an arbitrary input state.
Leaving aside the integration over the velocity distribution, we find that the cross-correlation for indistinguishable particles can be expressed as:
\begin{align}
	\Gtwo_{cd} \big\rvert_\text{indisc.}
	& = \frac{1}{4} \left( \Gtwo_{aa} + \Gtwo_{bb} + \Delta \right) \; , \quad \Delta = 2 \eta^2 \Real\!\left[ e^{2i\phi} \avg{\hat a^\dagger_{\mathbf v_a} \hat a^\dagger_{\mathbf v_a} \hat a_{\mathbf v_b} \hat a_{\mathbf v_b}} \right] \; , \\
	\intertext{whereas that of distinguishable particles reads}
	\Gtwo_{cd} \big\rvert_\text{disc.}
	& = \frac{1}{4} \left( \Gtwo_{aa} + \Gtwo_{bb} + 2 \Gtwo_{ab} \right) \; .
\end{align}
Here, the correlators appearing in the right-hand side are taken at time $t_1$, that is immediately after the the atom beams have been produced.
The term $\Delta$ corresponds to an interference between single-particle matter waves.
It depends on both the relative phase between the atom beams and the relative phase between the laser beams used for Bragg diffraction.
The latter is counted once for the atomic mirror and once for the atomic beam-splitter.
Twin beams with perfect correlations in their population would have a fully random relative phase. In our experiment however, the population imbalance between the atom beams could entail a residual phase coherence.
Instead, the relative phase between the laser beams was left uncontrolled and its value was randomly distributed between two repetitions of the experiment.
As a result, the term $\Delta$ must average to zero and the visibility of the HOM dip be given by Eq.~\ref{eq:visibility_bound}, as observed in the experiment.
Following Ref.~\cite{Lewis-Swan2014}, we also note that Eq.~\ref{eq:visibility_bound}  entails the ultimate bound for waves interfering on the beam-splitter: because waves must fulfil the Cauchy–Schwarz inequality, $\Gtwo_{ab} < \sqrt{\Gtwo_{aa} \Gtwo_{bb}}$, the visibility of the classical dip cannot exceed 0.5.

The above results holds true for a finite integration over the atomic velocity distribution if two conditions are met:
(i) It must remain impossible to distinguish the atoms entering the beam-splitter through channel $a$ from the atoms entering through channel $b$ once they have exited the beam-splitter;
(ii) The transformation matrix of the beam-splitter must keep the same form after integration.
In our experiment, the second condition is naturally satisfied because the Bragg diffraction only couples atomic states with a well defined momentum difference and we fulfil the first condition by reducing the integration volume as much as it is necessary.

\clearpage
\renewcommand\thefigure{S\arabic{figure}} 
\setcounter{figure}{0}  
\renewcommand\theequation{S\arabic{equation}} 
\setcounter{equation}{0}  
\newpage

\begin{center}
{\LARGE{\textbf{Supplementary material}}}
\end{center}
\subsection*{Optimization of the Hong-Ou-Mandel dip}

\begin{figure}[!h]
	\centering
	\includegraphics[width=0.7\columnwidth]{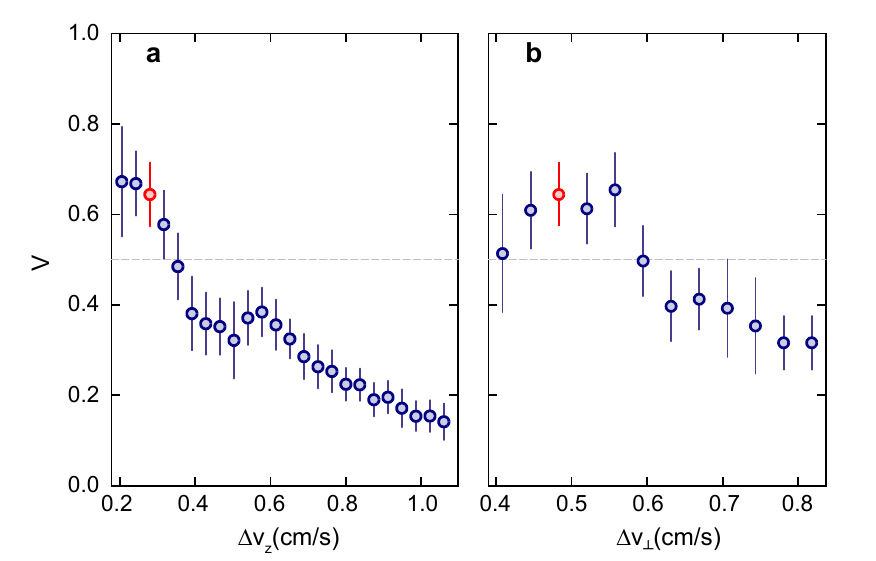}
	\caption{{\bf Visibility as a function of the integration volume.} 
	\textbf{a,}
	Visibility as a function of the longitudinal integration interval $\Delta v_{z}$. The transverse integration interval is kept constant at $\Delta v_{\perp} = \SI{0.48}{\cm \per \s}$ .
	\textbf{b,}
	Visibility as a function of the transverse integration interval $\Delta v_{\perp}$. The longitudinal integration interval is kept constant at $\Delta v_{z} =\SI{0.28}{\cm \per \s}$. The red points mark the values discussed in the main text.
	}
	\label{suppfig:1}
\end{figure}

The visibility of the Hong-Ou-Mandel dip is plotted in Fig.~\ref{suppfig:1} as a function of the longitudinal (a) and transverse (b) integration volume. The red dots identify the integration volume used in Fig.~\ref{fig:3} of the main text and correspond to a compromise between signal to noise ratio and visibility amplitude.
As we shrink the integration volumes, the dip visibility first increases and then reaches a saturation value, meaning that the integration volume becomes smaller than the elementary atomic modes \supercite{Rarity1989, Treps2005, Morizur2011}. Reducing further the integration volume only leads to an increase of the statistical uncertainty.

The visibility $V$ is obtained by fitting the cross-correlation function $\Gtwo_{cd}(\tau)$ measured in the experiment with the empirical function:
\begin{align*}
f(\tau) = \Gtwo_{\text{bg}} \Big( 1 - V \exp\big(-(\tau-\tau_{0})^2/2 \sigma^2\big) \Big) \; ,
\end{align*}
where the background correlation $\Gtwo_{\text{bg}}$, the center of the dip $\tau_{0}$ and the width of the dip $\sigma$ are all left as free parameters.

\newpage
\subsection*{Stability of the atom number in the output ports}

\begin{figure}[!h]
	\centering
	\includegraphics[width=0.7\columnwidth]{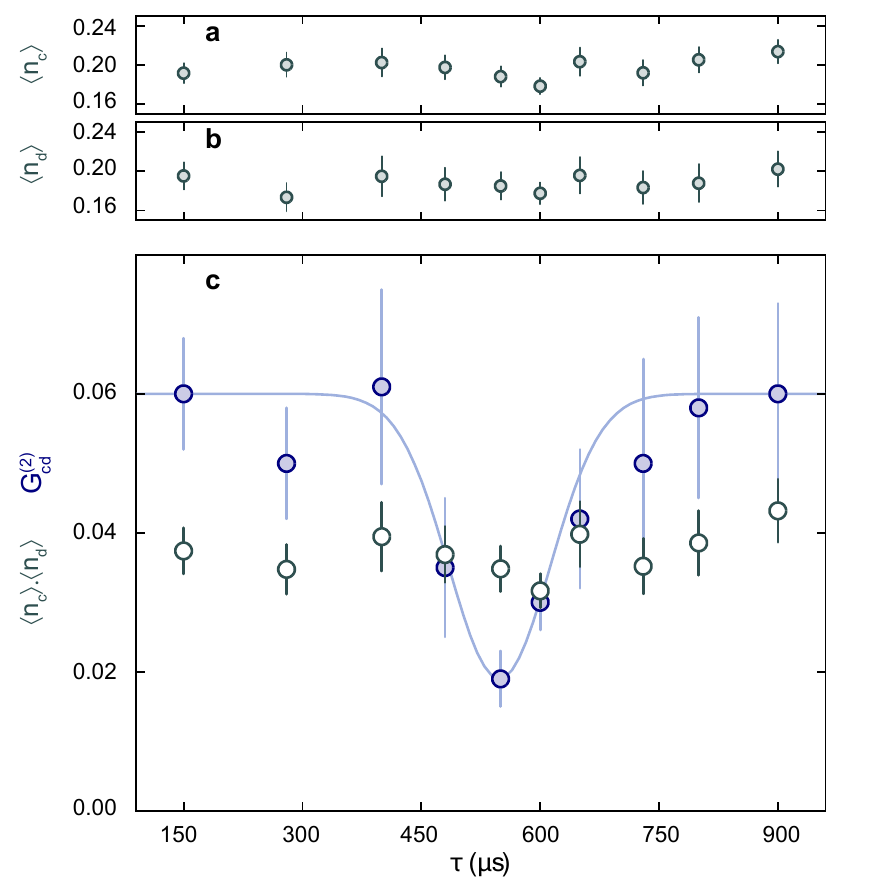}
	\caption{{\bf Stability of the output population over propagation duration.} 
	\textbf{a,}
	Averaged atom number detected in $\mathcal{V}_{c}$ as a function of the propagation time $\tau$. The mean value of $n_{c}$ is $\num{0.20}$ with a standard deviation of $\num{0.01}$.
	\textbf{b,}
	Averaged atom number detected in $\mathcal{V}_{d}$ as a function of the propagation time $\tau$. The mean value of $n_{d}$ is $\num{0.19}$ with a standard deviation of $\num{0.01}$.
	\textbf{c,}
	The cross-correlation between the output ports $c$ and $d$ (solid blue circles), corresponding to the HOM dip, is compared to the product of the average density populations $\langle n_{c} \rangle \! \cdot \! \langle n_{d}\rangle$ (open gray circles). The product of the averaged population is constant while the cross correlation exhibits a dip around $\tau = \SI{550}{\us}$.
	}
	\label{suppfig:2}
\end{figure}

The mean detected atom number in the output ports $c$ and $d$ is plotted as function of $\tau$ in Fig.~\ref{suppfig:2}. The mean atom number is constant as function of $\tau$ within the statistical uncertainty. To easily compare the atom number fluctuations to the variation of the cross-correlation across the HOM dip, the product of the averaged populations $\langle n_{c}\rangle \! \cdot \! \langle n_{d} \rangle$ and the cross-correlation $\Gtwo_{cd}$ are displayed together as a function of $\tau$ in Fig.~\ref{suppfig:2}c. In contrast to the cross-correlation, it is impossible to identify a marked reduction of $\langle n_{c}\rangle \! \cdot \! \langle n_{d} \rangle$ around $\tau = \SI{550}{\us}$. This confirms our interpretation of the dip as a destructive two-particle interference.

% -----------
% References
% -----------
\clearpage
\newpage
\printbibliography

\end{doublespace}
\end{document}